\documentclass[journal]{IEEEtran}
\usepackage{amsmath, amsfonts, amssymb}
\usepackage{algorithmic}
\usepackage{algorithm}
\usepackage{array}
\usepackage[caption=false,font=normalsize,labelfont=sf,textfont=sf]{subfig}
\usepackage{textcomp}
\usepackage{stfloats}
\usepackage{url}
\usepackage{verbatim}
\usepackage{graphicx}
\usepackage{cite}
\usepackage{siunitx}
\usepackage{xcolor}

\def\inserteq#1#2{\begin{equation}{#1}\label{#2}\end{equation}} 
\begin{document}

\title{Characterization of Silicon-Membrane TES Microcalorimeters for Large-Format X-ray Spectrometers with Integrated Microwave SQUID Readout}

\author{
Avirup Roy, Robinjeet Singh, Joel C. Weber, W. B. Doriese, Johnathon Gard, 
Mark W. Keller, John A. B. Mates, Kelsey M. Morgan, Nathan J. Ortiz, 
Daniel S. Swetz, Daniel R. Schmidt, Joel N. Ullom, Evan P. Jahrman, 
Thomas C. Allison, Sasawat Jamnuch, John Vinson, Charles J. Titus, 
Cherno Jaye, Daniel A. Fischer, Galen C. O’Neil

\thanks{A. Roy (avirup.roy@nist.gov), R. Singh, J. C. Weber, and J. Gard are with the Department of Physics, University of Colorado Boulder, Boulder, CO 80309 USA.}
\thanks{W. B. Doriese, M. W. Keller, J. A. B. Mates, K. M. Morgan, N. J. Ortiz, D. S. Swetz, D. R. Schmidt, J. N. Ullom, and G. C. O'Neil are with the National Institute of Standards and Technology, Boulder, CO 80305 USA.}
\thanks{E. P. Jahrman, T. C. Allison, S. Jamnuch, J. Vinson, C. J. Titus, C. Jaye, and D. A. Fischer are with the National Institute of Standards and Technology, Gaithersburg, MD 20899 USA.}
 \thanks{This work was performed under financial assistance award 70NANB23H027 from the U.S. Department of Commerce, National Institute of Standards and Technology.}
}
\markboth{IEEE TRANSACTIONS ON APPLIED SUPERCONDUCTIVITY, VOL. XX, NO. XX, MONTH 202X}{}
\maketitle
\IEEEpubidadjcol

\begin{abstract}

We present the electro-thermal characterization of transition-edge sensor (TES) detectors suspended on Si membranes fabricated using a silicon-on-insulator (SOI) wafer. The use of an all-silicon fabrication platform, in contrast to the more commonly used silicon nitride membranes, is compatible with monolithic fabrication of integrated TES and SQUID circuits. The all-silicon architecture additionally allows efficient use of focal plane area; the readout circuitry may be positioned out of the focal plane by bending a thinned portion of the chip. Compatibility with integrated fabrication and efficient use of focal plane area provide a path to an efficient soft X-ray spectrometer. 

This work is motivated by our goal to develop a 10,000-pixel TES spectrometer to overcome critical measurement limitations in catalysis research. The characterization of fragile, carbon-based intermediates via techniques like Resonant Inelastic X-ray Scattering (RIXS) is often precluded by the slow, high-flux nature of existing technologies. The new instrument will allow for fast RIXS measurements to be made without causing sample damage. We verify the detector models and measure the energy resolution using a pulsed optical laser, demonstrating the viability of this approach for the final instrument to be deployed at the National Synchrotron Light Source II (NSLS-II).
 
\end{abstract}


\section{Introduction}
\IEEEPARstart{L}{ow}-temperature X-ray microcalorimeters offer a powerful combination of high energy resolution and high detection efficiency, making them ideal for chemical analysis~\cite{barba2024formal,barba2025following, liu2024vitreous, bhowmick2025active}. We are developing a 10,000-pixel soft X-ray spectrometer based on this technology for deployment at the X-ray beamline 7-ID-1, which is operated by the National Institute of Standards and Technology (NIST) at the National Synchrotron Light Source II (NSLS-II). A primary scientific driver for this instrument is rapid Resonant Inelastic X-ray Scattering (RIXS) measurements for \textit{operando} catalysis studies. For background, RIXS provides detailed insight into a material’s electronic structure\cite{de2024resonant} with several applications in chemistry\cite{lundberg2019resonant}, including the quantification of key electronic parameters in catalysis~\cite{cui2017revealing, liu2024manipulating}.

The proposed array provides distinct advantages for RIXS measurements of catalysts. Specifically, energy-dispersive microcalorimeters allow for the use of large X-ray spot sizes, enabling the study of radiation-sensitive compounds, such as carbon-based intermediates~\cite{titus2017edge}. Additionally, the large throughput provided by the 10,000-pixel array will enable rapid RIXS measurements for time-resolved studies. Finally, given that the depth of coordination chemistry insights is closely tied to spectral performance~\cite{van2012multispectroscopic}, the target energy resolution of \SI{0.3}{eV} FWHM below \SI{300}{eV} is critical for facilitating the unambiguous identification of carbon-based species and probing ligand-metal interactions at active sites.

\begin{figure}
    \centering
    \includegraphics[width=1\linewidth]{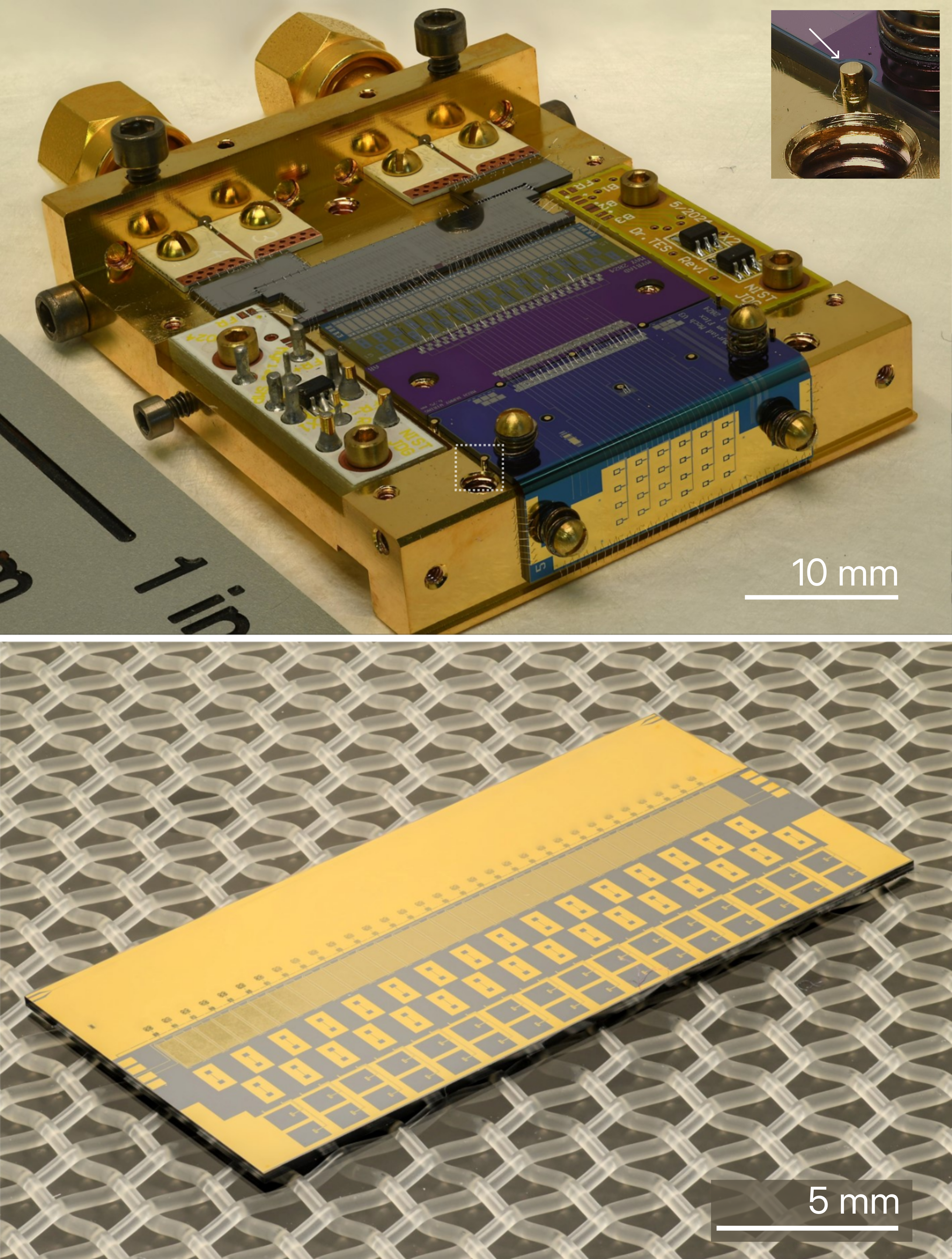}
    \caption{(Top) A focused combined photograph of detector box containing a 24-pixel TES array bent 90 degrees around the corner using a a section of $\sim$~\SI{4}{\micro m} thick flexible silicon. The TES detector array is wire-bonded to an inductor chip, a shunt chip and finally a microwave multiplexer chip that is instrumented using a single coaxial feedline. Copper ``ground bosses'' are machined from the detector cold plate for RF cavity mode suppression (dotted white square, and inset). (Bottom) A 24-pixel integrated chip, showing the TES, inductors, shunts and SQUIDs fabricated on the same die in a 19+ layer process~\cite{singh2025lithographic}. }
    \label{fig:combined-stackshot}
\end{figure}

We base our detector design on transition-edge sensors (TESs), a mature and scalable microcalorimeter technology~\cite{ doriese2017practical, bandler2019lynx, smith2024development}. The thermodynamic limit to the energy resolution ($\Delta E$) in TESs is of the form:
\inserteq{\Delta E \propto 2.355\sqrt{k_\mathrm{B}T_\mathrm{c}^2 C \sqrt{(1 + 2\beta_\mathrm{I})(1+M^2)}/\alpha_\mathrm{I}},}{}
 where $T_\mathrm{c}$ is the critical temperature, $C$ is the total heat capacity of the detector, $\alpha_\mathrm{I} \equiv T/R \cdot \partial R/\partial T$ is the temperature sensitivity, $\beta_\mathrm{I} \equiv I/R \cdot \partial R/\partial I$ is the current sensitivity~\cite{irwin2005transition}, and  $M$ denotes the ``excess" Johnson noise term. 
 Achieving our target resolution thus requires developing TESs with a low $T_\mathrm{c}$ ($\approx$ \SI{30}{mK}). Simultaneously, achieving a large collecting area requires a scalable architecture. Our approach combines two key technologies: integrated fabrication of TES and SQUID components~\cite{singh2025lithographic} and the use of a flexible silicon layer to create a compact spectrometer focal plane~\cite{o2024flexible} (see Fig.~\ref{fig:combined-stackshot}). A key requirement for both fabrication strategies is the use of a silicon substrate, making silicon-on-insulator (SOI) wafers the necessary starting material. Consequently, our TESs must be supported by silicon membranes, in contrast to more traditional designs on silicon nitride. We avoid nitride membranes because etching it away to clear the substrate for the readout circuitry would worsen the pristine substrate required for high-$Q$ superconducting resonators.
 
 Therefore, a complete electro-thermal characterization of these silicon-based devices is a critical step. In this work, we present such a model for our TESs on SOI, developed using I-V curves, complex admittance, pulse response, and noise spectra. We find that the silicon membranes provide acceptable control of thermal conductance and that the devices exhibit reasonably sharp superconducting transitions. However, our model also reveals an unexplained excess heat capacity that is not present in similar devices fabricated on traditional silicon nitride membranes.
 We validate this model using spectral results from a pulsed green laser ($\langle \lambda \rangle \approx \SI{515}{nm}$). This technique allows us to directly measure the energy resolution, and we find good agreement between the achieved resolution and that predicted by our model. We use the laser data to also study the detector's calibration curve and characterize the thermal crosstalk effects arising from the flood illumination of the array. This optical photon-counting technique is a powerful tool during detector development~\cite{hokin2014narrow, jaeckel2019energy, roy2024pulsed} and deployment, as all conveniently available soft X-ray fluorescence lines are too broad for accurately characterizing sub-eV resolution.

\section{Experiment}
We have investigated two different microchips in this work, hereafter referred to as the “flex chip” and the “solid chip.” Both chips were fabricated on the same silicon-on-insulator (SOI) wafer and contain a variety of nominally identical TES devices.

The distinction arises from the application of the deep reactive-ion etching (DRIE) step used to release the TES membranes. On the solid chip, this etch only defines the membranes. On the flex chip, this same DRIE step is extended to also remove the underlying handle wafer in a specific region. This creates a $\sim$ \SI{1.5}{mm} long, $\sim$ \SI{4}{\micro m} thick flexible section of device-layer silicon that carries the wiring to the detectors and allows the array to be positioned out-of-plane from the readout circuitry.

The flex chip was measured in the sample box shown in Fig.~\ref{fig:combined-stackshot}, whereas the solid chip was measured in a separate enclosure. Both setups used interface chips with the same target values for inductance and shunt resistance.

\begin{figure}
    \centering
    \includegraphics[width=\linewidth]{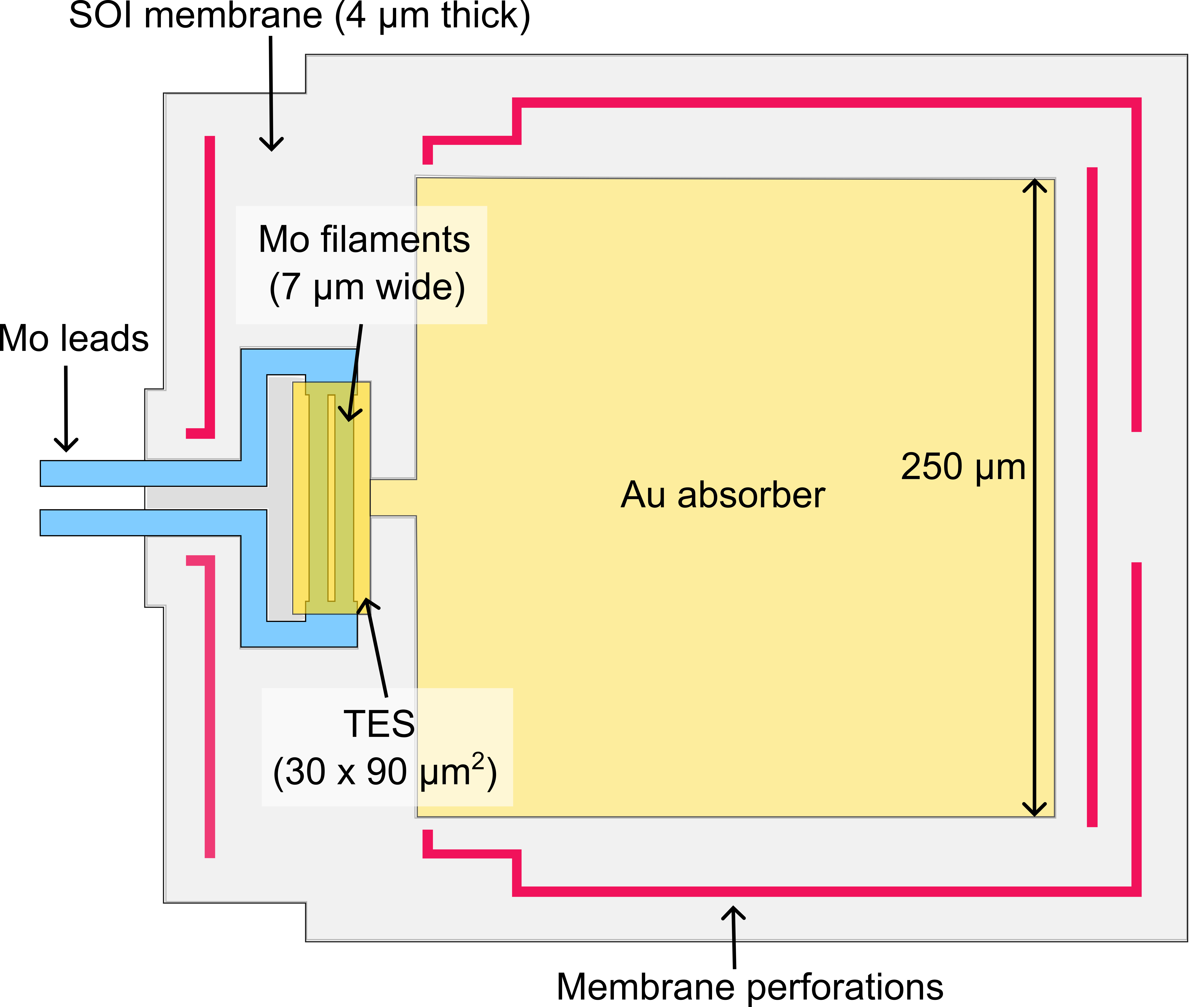}
    \caption{Schematic representation of a single TES detector. The Mo film (blue) is wet-etched to create two \SI{7}{\micro m} wide filaments that are used to define $T_\mathrm{c}$. The sensor and its ``sidecar'' absorber are suspended on a $\sim$ \SI{4}{\micro m} thick silicon membrane (grey). Membrane perforations (red) are used to define the thermal conductance of the device to the bath. The jagged outermost boundary represents the outline of back etch which defines the Si membrane.}
    \label{fig:single-pixel}
\end{figure}

The detector chips contain a 4\,$\times$\,6 array of TES pixels. While the Mo\,/\,Au sensors share uniform dimensions of \SI{30}{\micro m}\,$\times$\,\SI{90}{\micro m} and thicknesses of \SI{46}{nm}\,/\,\SI{630}{nm}, the support-leg geometries vary across the array. This design allows us to target a range of thermal conductance ($G$) values on a single chip for characterization. The sensors are fabricated using the ``hasTES'' process~\cite{weber2020development} with the Mo layer etched into \SI{7}{\micro m} wide filaments before Au deposition. The filaments enable a useful level of lithographic control of $T_\mathrm{c}$ and $\alpha_\mathrm{I}$.
This narrow rectangular sensor design has been shown to reduce magnetic field sensitivity~\cite{wakeham2023refinement}. The sensor is thermally coupled to a \SI{250}{nm} thick, \SI{250}{\micro m} square gold ``sidecar'' absorber~\cite{alpert2019high} with a \SI{250}{nm} thick Au link, measuring \SI{18}{\micro m}\,$\times$\,\SI{15}{\micro m}. The entire TES and absorber assembly is suspended on a silicon membrane that is $\sim$~\SI{4}{\micro m} thick. The Si membrane has a $\sim \SI{350}{nm}$ thick SiO$_2$ layer deposited with plasma-enhanced chemical vapor deposition to prevent Mo-Si inter-diffusion. The normal resistance ($R_{\rm n}$) of the TESs is about \SI{7}{\milli \ohm}, and the 3-square design is a good match to our bias circuit. The TES schematic is shown in Fig. \ref{fig:single-pixel}.

The detector chip is wire-bonded to interface chips that provide a shunt resistance $R_\mathrm{shunt}$ $\approx $ \SI{250}{\micro \ohm} and an inductance $L \approx$ \SI{120}{nH}. The interface chips are connected to a microwave multiplexer ($\mu$mux) chip capable of reading out 31 resonators placed in a \SI{500}{MHz} wide band~\cite{mates2017simultaneous}. A single RF coaxial line carries the multiplexed signal from the $\mu$mux feedline inside the detector box. As it leaves the cryostat, the signal passes through two amplification stages. The first is a High-Electron-Mobility Transistor located at the \SI{3}{K} stage, followed by a low-noise amplifier at the \SI{40}{K} stage.

To mitigate RF cavity modes and preserve the high microwave resonator quality factors ($Q_i > 100,000$) needed for large multiplexing factors ($\approx$ 500), we have incorporated a unique  design feature into the detector box. We have machined the cold plate to form cylindrical copper ``ground bosses'', approximately \SI{450}{\micro m} tall and \SI{400}{\micro m} wide, positioned on a \SI{5}{mm} pitch within the region designated for resonator fabrication. We utilize the same DRIE step that is already part of our fabrication process to define membranes and the chip outline to also define clearance holes in the silicon chips, accommodating these bosses. The top panel in Figure \ref{fig:combined-stackshot} shows a test package that validates this mechanical design. Although the chip shown uses this region for a wiring passthrough, the final instrumented chips will have their microwave resonators fabricated in this bossed area.

The detector box is mounted to the \SI{10}{mK} stage of a dilution refrigerator and enclosed in two nested layers of high-permeability cryogenic magnetic shielding; we have measured a residual field $\lesssim$ \SI{50}{nT} normal to the plane of the detectors below \SI{1}{K} using a cryogenic flux-gate magnetometer in prior cooldowns which was placed in the same magnetic environment as the detector box. A fiber-coupled pulsed-laser system provides flood illumination of the detector array with photons with mean wavelength around \SI{515}{nm} ($\approx$ \SI{2.4}{eV}).  The light is delivered by a \SI{200}{\micro m} core, multimode bare glass optical fiber, which is coupled using an SMA connector to a light-tight housing on top of the detector box. Inside this housing, a collimating mirror directs the photons onto the detector array in a spot that is $\sim \SI{12}{mm}$ wide. The detector array is covered by a silicon aperture chip which has 24 DRIE-etched \SI{150}{\micro m} square holes centered $\sim \SI{45}{\micro m}$ above every absorber in the array. About \SI{100}{nm} of Au is sputtered on the aperture chip to provide heatsinking to the cold plate using $\sim$ 30  gold wire bonds.

\section{Results}


\subsection{Thermal conductance}



\begin{figure}
    \centering
    \includegraphics[width=1\linewidth]{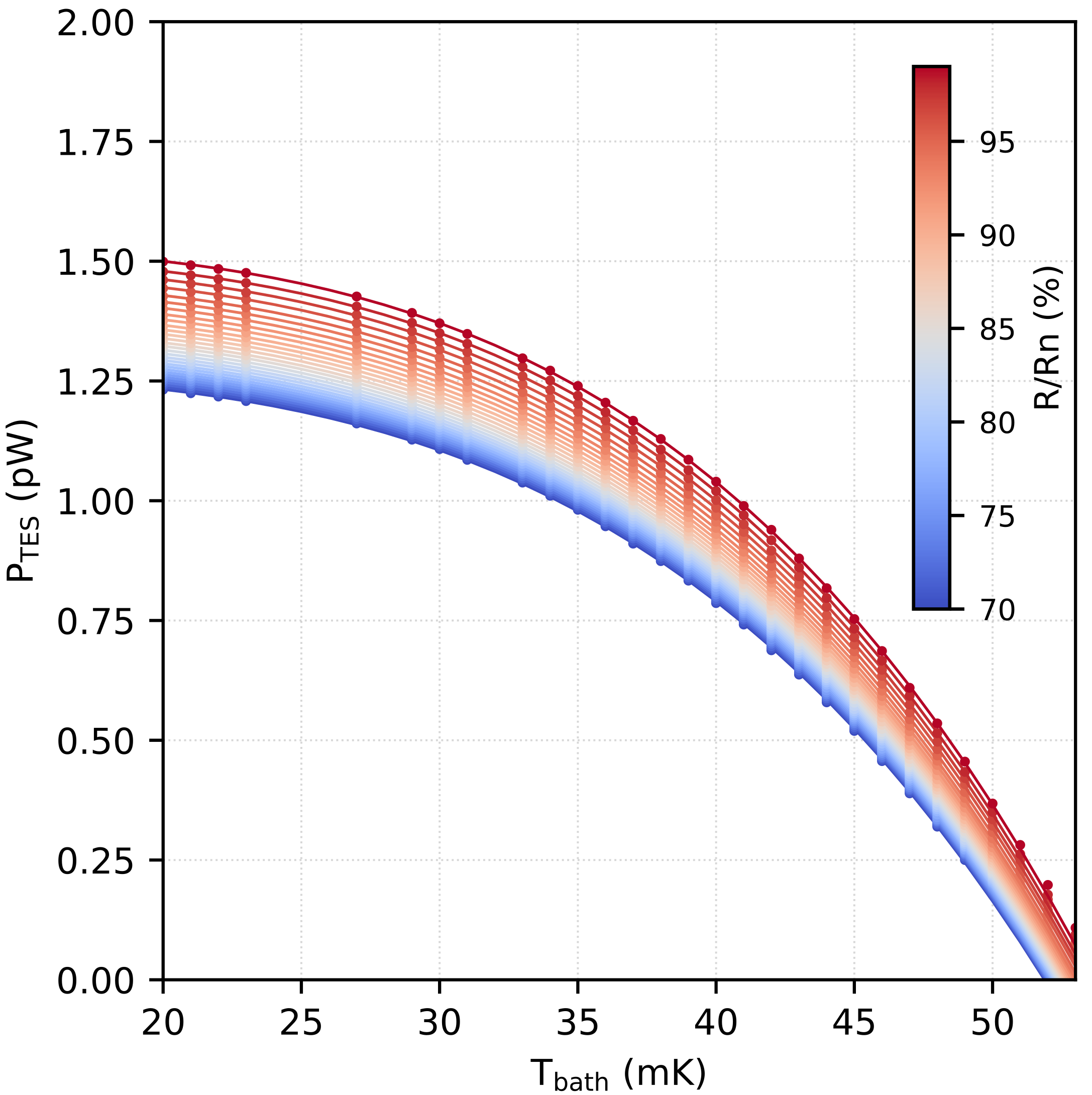}
    \caption{Power-law fits to the bias power vs. bath temperature for a single representative pixel at bias points high in the transition where the current dependence of the resistance is vanishingly small. Fit results at 80\% $R_n$: $G_\mathrm{\SI{30}{mK}} = \SI{21}{pW/K}$, $n = 3.8$, and $T_\mathrm{c} = \SI{53}{mK}$.}
    \label{fig:PvsTbase}
\end{figure}

We measured the thermal conductance, $G$, for a set of TES devices on the solid chip, which were designed with varying support-leg geometries to achieve different $G$-values. The measurement involved acquiring I-V curves at bath temperatures ($T_\mathrm{bath}$) ranging from \SI{20}{mK} to \SI{60}{mK}. For each device, the thermal conductance was extracted by fitting the relationship between TES bias power, $P_\mathrm{TES}$, and bath temperature to the power-law model shown in Fig.~\ref{fig:PvsTbase}:
\inserteq{P_\mathrm{TES} = k(T_\mathrm{TES}^n - T_\mathrm{bath}^n) \Theta(T_\mathrm{TES} - T_\mathrm{bath}),}{eq1}
where $k$ is a thermal conductivity coefficient, $T_\mathrm{TES} \approx T_\mathrm{c}$ is the operating temperature of the TES, $n$ is the power-law exponent, and $\Theta$ is the Heaviside step function.

\begin{figure}
    \centering
    \includegraphics[width=0.9\linewidth]{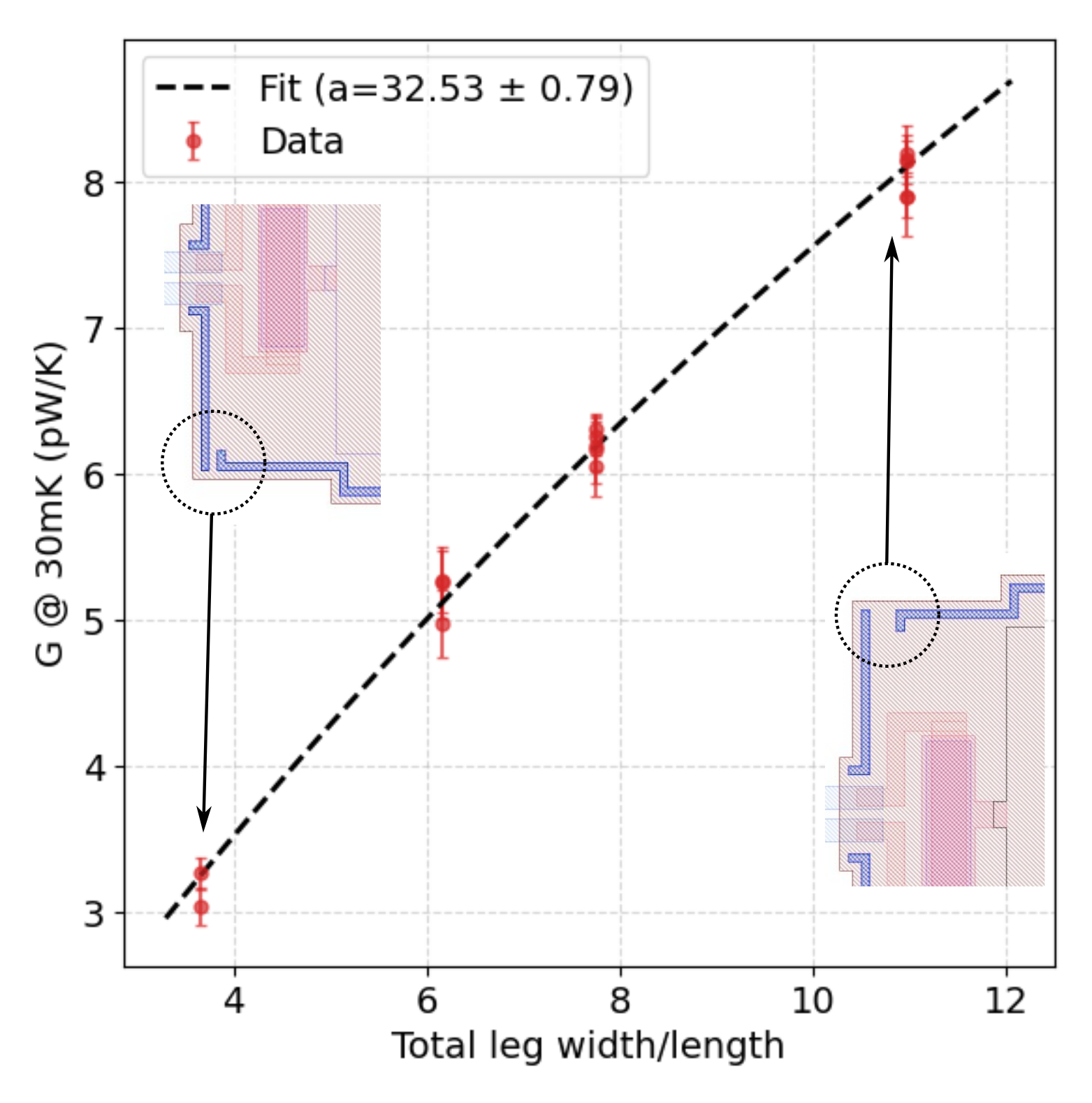}
    \caption{Summary plot of thermal conductance vs. leg aspect ratio ($\Sigma w/l$) for 10 TES detectors. The data are fit with a single free parameter $a$, representing the thermal conductance of bulk of the Si membrane, which is connected in series with the calculable link conductances to the bath. 
    }
    \label{fig:Gvsaspect}
\end{figure}

An advantage of working with Si membranes is that the phonon contribution to the thermal conductance can be calculated directly from first principles. The Debye model predicts that 
\inserteq{G = \frac{\gamma T^3 c \lambda}{3}\frac{A}{l}, }{eq2}
where $\gamma$ = 0.57~JK$^{-4}$m$^{-3}$, $\gamma T^3$ is the phonon contribution to the heat capacity, $\lambda = \sqrt{2}t$ is the phonon mean free path for a membrane with thickness $t$, $c$ is the speed of sound in silicon ($\approx$ \SI{8520}{m/s}), and $A$ and $l$ are the area and length of the thermal link, respectively. In this framework, we can fit the thermal conductance of all pixels in the array against a dimensionless parameter representing the sum of the ratios of the widths to the lengths of the links from the TES to the bath. This is shown in Fig. \ref{fig:Gvsaspect}, where the data are fit with the function
\inserteq{G = \frac{\gamma T^3 c \lambda t}{3} \left[\left(\sum_i\frac{w_i}{l_i}\right)^{-1} + \frac{1}{a} \right]^{-1}, }{eq3}
where $w_i$ and $l_i$ are the leg widths and lengths respectively, and $a$ is the single fit parameter, representing the thermal conductance of the internal bulk of the Si membrane that is in series with the legs to the bath.
We find that all pixels in the array have transition temperatures between \SI{53}{mK} and \SI{55}{mK}. 

\subsection{Complex Admittance}
The dynamic response of the TES is characterized by measuring its complex admittance; this allows us to derive $\alpha_\mathrm{I}$, $\beta_\mathrm{I}$, $C$ and $G$. The admittance data shown here were obtained from TESs on the flex chip.  For this, we applied a multitone voltage bias ($V_\mathrm{exc}(t) = V_\mathrm{DC} + A \sum_i \mathrm{cos}(\omega_i t)$) at $>$ 20 frequencies ($\omega_i$) in the \SI{5}{Hz} - \SI{50}{kHz} range  and recorded the detector's current response $I(t)$.

When the TES is superconducting, the circuit's physical response is a simple single-pole L-R network. We therefore model the frequency response in the superconducting state as
\inserteq{I_\mathrm{LR}(\omega) = \frac{I_0}{1 + j \omega \tau_\mathrm{el}} e^{j\omega \Delta t.}}{eq:timedelay} where $\tau_\mathrm{el} = L/R_\mathrm{shunt}$ is the time constant, $L$ is the inductance, $R_\mathrm{shunt}$ is the shunt resistance, and $\Delta t$ is a time offset between the excitation and readout systems. We fit the model to $I_\mathrm{LR}(\omega)$, the Fast Fourier Transform of $I(t)$ measured with the TES in the superconducting state. We decouple the fit parameters in a two-step process. First, we fit the magnitude of the frequency response, $|I_\mathrm{LR}(\omega)|$, which is independent of $\Delta t$, to determine the DC response term $I_0$ and the electrical time constant $\tau_\mathrm{el}$. Next, with $I_0$ and $\tau_\mathrm{el}$ determined, we fit the real part of the measured data to extract  $\Delta t$. For further analysis, all  TES measurements are multiplied by the correction factor $e^{-j \omega \Delta t}$ to recover the true complex response. We use the value $R_\mathrm{shunt}=\SI{250}{\micro\ohm}$ from four-wire measurements on other copies of the interface chip to find $L=\SI{120}{nH}$, the value well matched to our expectation from the on-chip inductance plus the inductance of the wiring.

The complex admittance of the TES, $Y_\mathrm{TES}$, is determined from the ratio of the measured biased and superconducting transfer functions, following the method outlined in Lindeman \textit{et al.}~\cite{lindeman2004impedance}:
\inserteq{Y_\mathrm{TES} = \left[\left(\frac{I_\mathrm{TES}(\omega)}{I_\mathrm{LR}(\omega)}-1\right)(R_\mathrm{shunt}+j\omega L)\right]^{-1}}{}
where $I_\mathrm{TES}$ is measured at multiple DC excitation values.
We fit this data using the closed-form solutions for electrothermal models provided by Maasilta~\cite{maasilta2012complex}. 

\begin{figure}
    \centering
    \includegraphics[width=1\linewidth]{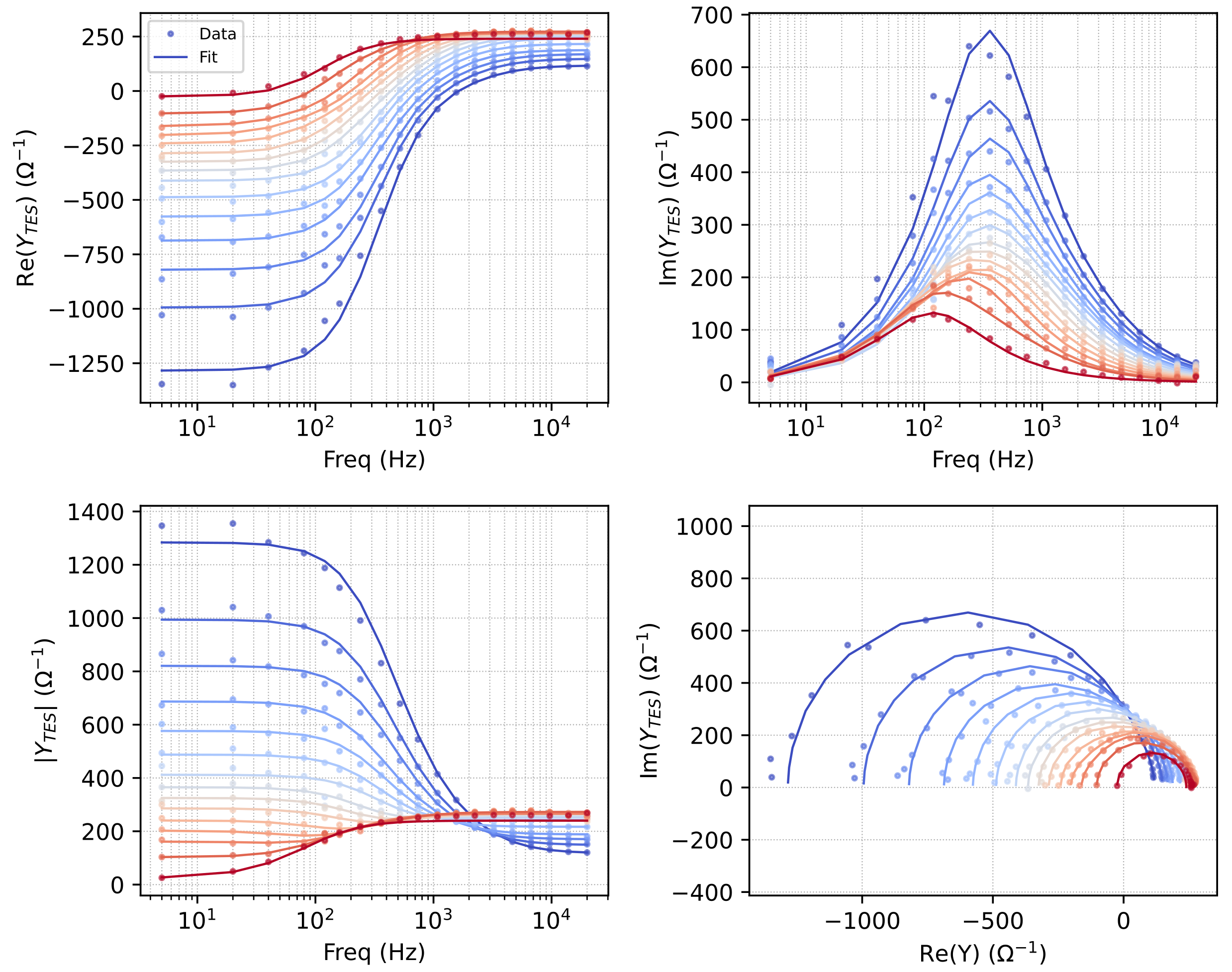}
    \caption{Complex admittance plots for a single pixel taken at bias points between 10\% and 90\% $R_n$ (blue to red). The magnitude, real, imaginary parts of $Y_\mathrm{TES}$ vs. excitation frequency are shown along with the Nyquist plot (clockwise from bottom left). Systematic deviations from model are observed below \SI{40}{Hz}. 
    }
    \label{fig:admittance}
\end{figure}

Figure~\ref{fig:admittance} shows the measured $Y_\mathrm{TES}$ for a single pixel at various bias points between 15\% and 90\% of normal resistance ($R_n$), with fits to the hanging heat capacity model. We found that a simple one-body model was insufficient to describe the data; it was necessary to use a two-body hanging heat capacity model, which indicates the presence of an additional heat capacity weakly coupled to the TES.  From these fits, we extract $\alpha_\mathrm{I}$, $\beta_\mathrm{I}$, $C_\mathrm{TES}$, $C_\mathrm{hanging}$, and $G_\mathrm{TES, hanging}$. We find that the hanging heat capacity accounts for $\sim$ 24\% of the total device heat capacity ($C_\mathrm{TES} + C_\mathrm{hanging}$). 

\begin{figure}
    \centering
    \includegraphics[width=\linewidth]{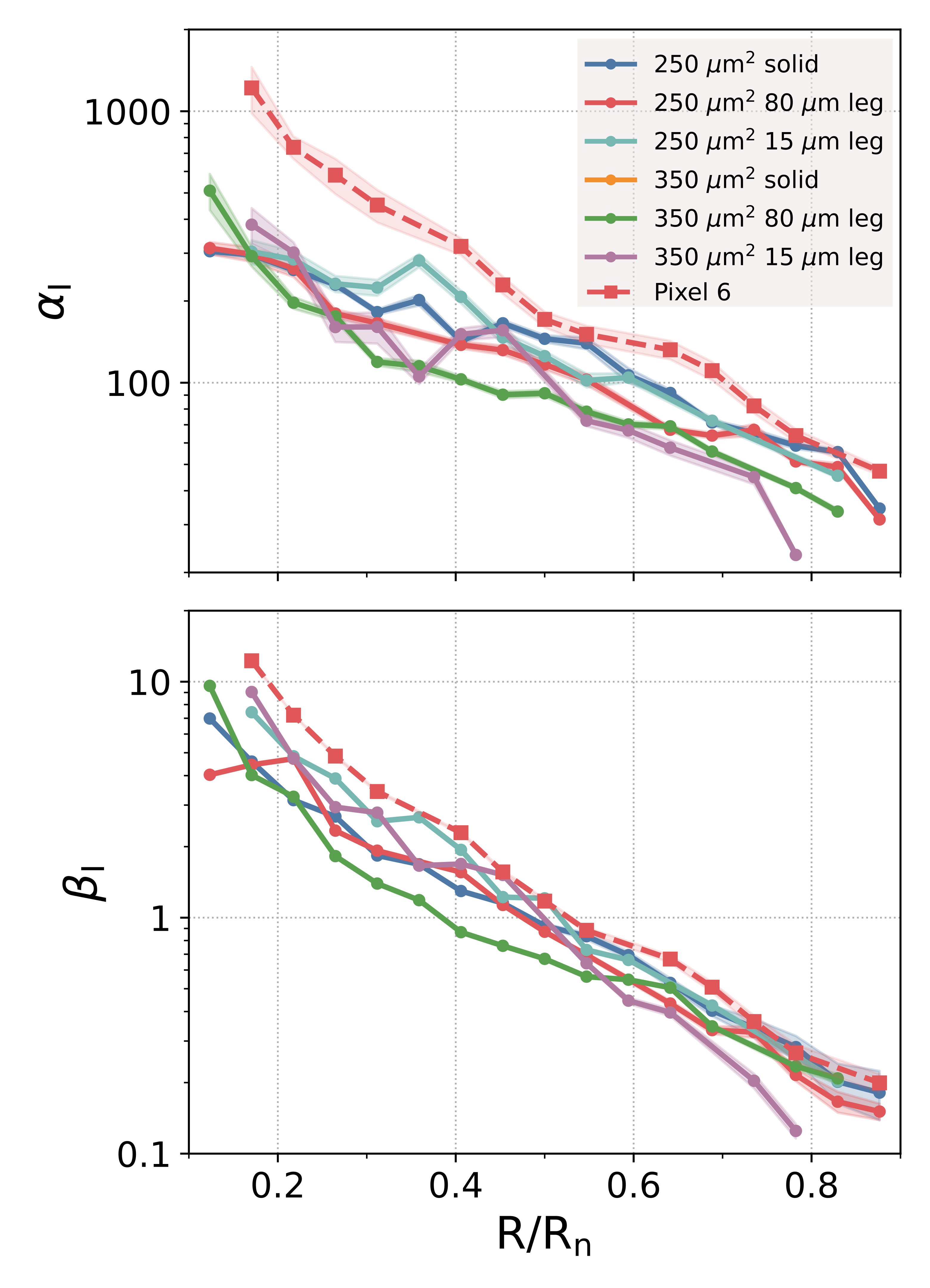}
    \caption{Weighted averages of transition parameters $\alpha_\mathrm{I}$ and $\beta_\mathrm{I}$ for fifteen devices shown as a function of bias point. There are three devices per design (solid circles), and a TES (Pixel~6: \SI{250}{\micro m} square absorber, \SI{80}{\micro m} wide leg) chosen for further analysis is shown separately (dashed squares). Shaded regions indicate the $1\sigma$ confidence intervals of the weighted means.}
    \label{fig:alpha-beta-vs-bias}
\end{figure}

The combined $\alpha_\mathrm{I}$ and $\beta_\mathrm{I}$ data from 15 devices are shown in Fig. \ref{fig:alpha-beta-vs-bias}. These devices vary only in absorber geometry and $G$, so we expect them to have similar transition parameters.

\begin{figure}
    \centering
    \includegraphics[width=\linewidth]{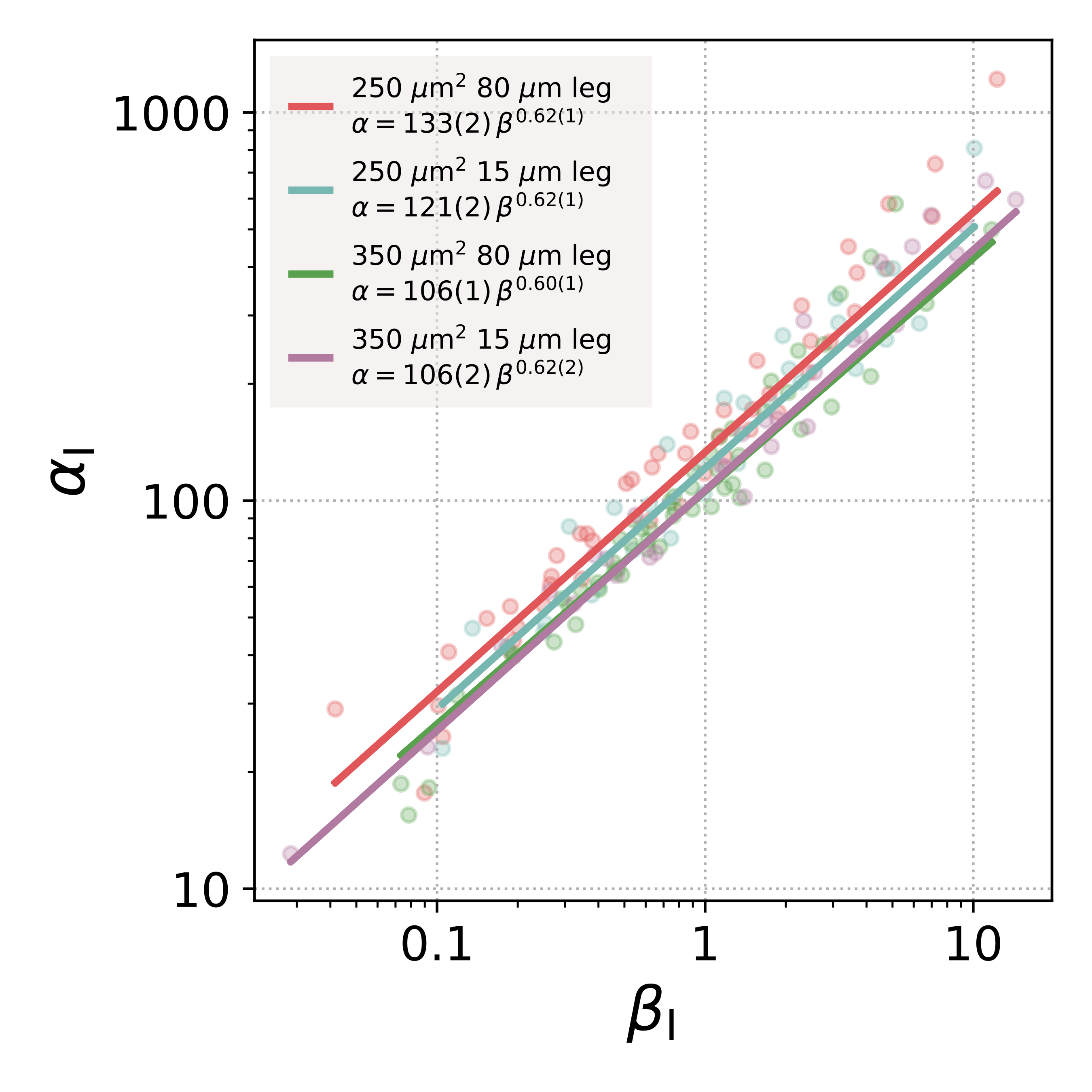}
    \caption{Power-law fits to weighted averages of $\alpha_\mathrm{I}$ and $\beta_\mathrm{I}$ parameters for twelve devices with four different designs---two different absorber sizes and two different thermal link geometries. $1\sigma$ uncertainties in the fit parameters are shown in parentheses on the last significant digits.}
    \label{fig:alpha-vs-beta}
\end{figure}

The relationship between $\alpha_\mathrm{I}$ and $\beta_\mathrm{I}$, shown in Fig. \ref{fig:alpha-vs-beta}, has been reported to follow the empirical power law $\alpha_\mathrm{I} = c \beta_\mathrm{I}^m$.  A fit to our data yields an exponent $m \approx 0.6$ for all the different absorber geometries and $G$ designs that we have included on the chip. This sub-linear behavior ($m<1$) contrasts with the nearly linear dependence ($m \approx 1$) observed in square TESs~\cite{jaeckel2023performance} but is consistent with results from other rectangular sensors. For example, similar Mo/Au devices showed $m \approx 0.6$~\cite{fabrega2022temperature}, and rectangular Ti/Au devices with 2 to 6 square designs exhibited $m \approx 0.7-0.8$, a behavior ascribed to weak-link effects~\cite{de2020high}.

\subsection{Noise and Pulse fit}

Noise and pulse data were taken on the solid chip. The average pulse is constructed from single-photon events generated by a \SI{2.4}{eV} laser diode driven by a pulsed current source outside the cryostat. By using the single-photon \SI{2.4}{eV} pulse, we are guaranteed to remain in the small-signal limit.

\begin{figure}
    \centering
    \includegraphics[width=1\linewidth]{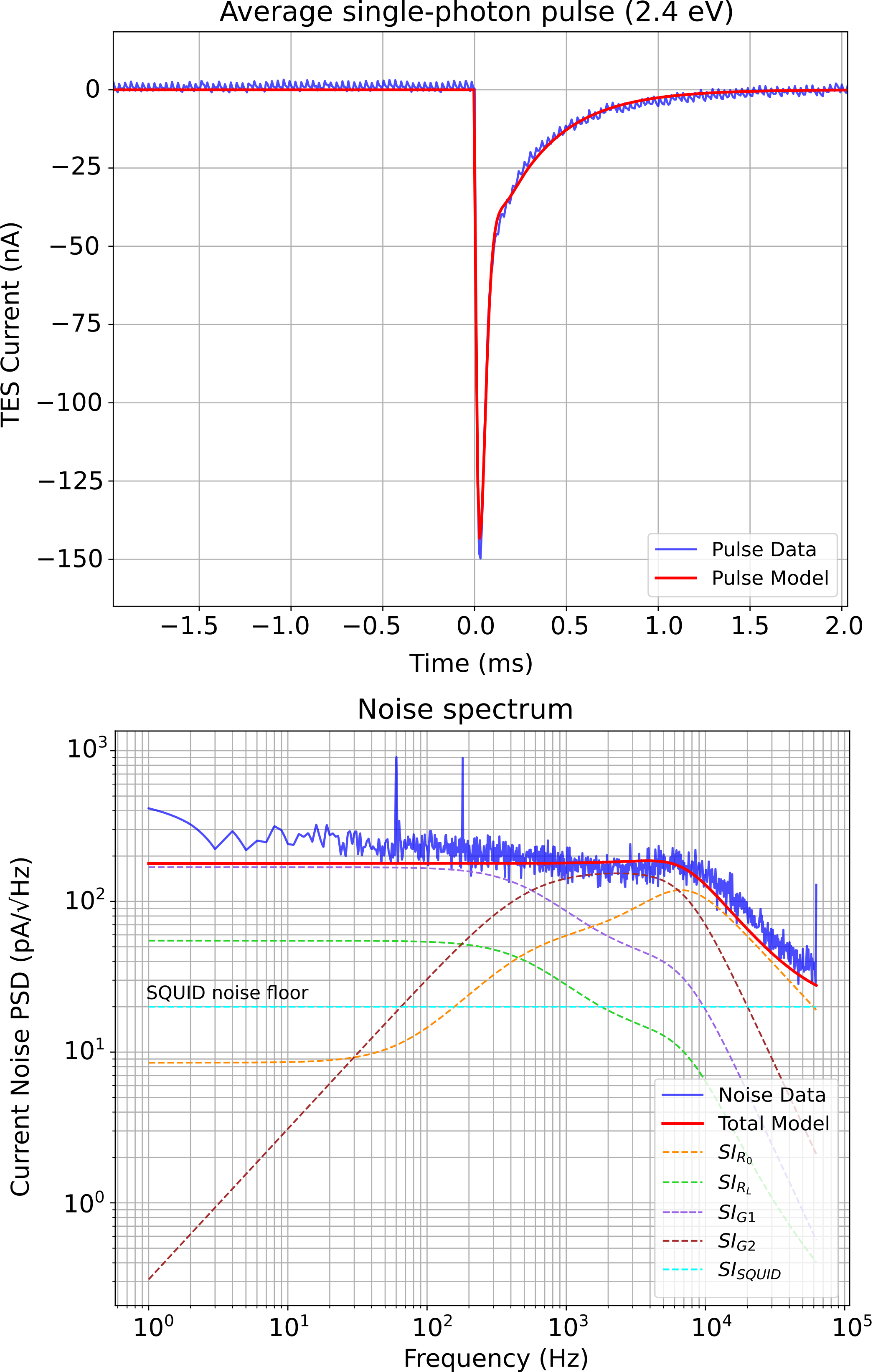}
    \caption{Combined pulse and noise fit for Pixel 6 at 7\% $R_n$. The pulse data is collected by averaging single-photon pulses from a \SI{2.4}{eV} laser diode. The high frequency ringing in the time trace is associated with the $\mu$mux readout. The SQUID white noise floor is at 20~pA/$\sqrt{\mathrm{Hz}}$. The excess Johnson noise term $M$ is included in the TES noise contribution $S_{I,R_0}$.}
    \label{fig:noise-pulse-fit}
\end{figure}

Analysis of the average pulse shape, shown in Fig. \ref{fig:noise-pulse-fit}, revealed two distinct decay time constants. This behavior is expected from the relatively large hanging heat capacity found in fits to $Y_\mathrm{TES}$. Therefore we use the same hanging heat capacity model to fit the pulse and noise data.

For the fit, we fixed  $R_\mathrm{shunt}$, $L$, $G_\mathrm{TES,bath}$ and the TES bias point parameters $R_0$ and $T_0$ based on the measurements in the previous sections. We treated the remaining terms in the thermal model as free parameters, namely $\alpha_\mathrm{I}$, $\beta_\mathrm{I}$, $C_\mathrm{TES}$, $C_\mathrm{hanging}$, the internal thermal conductance $G_\mathrm{TES,hanging}$, and an excess Johnson noise $M$. We used the values obtained from the admittance measurements on the nominally identical flex chip as initial guesses for the parameters; admittance data was collected at a higher bias point than the pulse data, so direct comparison of parameters between the chips was not possible. We then performed a simultaneous fit to both the average pulse shape and the noise spectrum.

Simultaneous fits of the noise and pulse taken for a detector biased at 7\% $R_{\rm n}$ are shown in Fig. \ref{fig:noise-pulse-fit}. All device parameters, including the best-fit values for the free parameters, are listed in Table \ref{tab:fitparams}. The model predicts a FWHM energy resolution of \SI{0.58}{eV}. Our model also shows that the presence of the second thermal body degrades the energy resolution by \SI{0.17}{eV}, or about 40\%. Because similar pixels on silicon nitride membranes show only a single decay constant, and the specific heat of crystalline silicon is negligible at \SI{50}{mK}, we attribute this excess heat capacity to the $\sim$ \SI{350}{nm} layer of PECVD SiO$_2$ on the membrane. Amorphous silica is known to exhibit anomalous excess heat capacity below \SI{1}{K} due to two-level systems~\cite{zeller1971thermal, phillips1972tunneling}. We plan to investigate this in the near term by selectively removing the oxide under the absorber. Ultimately, the implementation of overhanging absorbers~\cite{chervenak2004fabrication} will minimize the area of the membrane in contact with the detectors and significantly reduce this contribution.

\begin{table}[h!]
    \centering
    \caption{Parameters for the electrothermal model fit shown in Fig. \ref{fig:noise-pulse-fit}.}
    \label{tab:fitparams}
    \begin{tabular}{lccc}
        \hline
        \hline
        Parameter  & Value  \\
        \hline
        \multicolumn{2}{l}{\textit{Fixed Parameters}} \\
         $R_\mathrm{shunt}$ & \SI{250}{\micro \ohm} \\
         $R_0$ & \SI{470}{\micro \ohm} \\
         $L$ & \SI{120}{nH} \\
         $G_\mathrm{TES,bath}$ & \SI{93}{pW/K} \\
         $T_0$ & \SI{53}{mK} \\
         $T_\mathrm{bath}$ & \SI{21}{mK}\\
         $S_\mathrm{I, SQUID}$ & $\SI{20}{pA/\sqrt{Hz}}$ \\
         $E_\mathrm{init}$ & \SI{2.4}{eV}\\
         \hline
         \multicolumn{2}{l}{\textit{Fit Parameters}} \\
         $\alpha_\mathrm{I}$ & $1365 \pm 27$  \\
         $\beta_\mathrm{I}$ & $64 \pm 1$  \\
         $C_\mathrm{TES}$ & $161 \pm 2$  fJ/K \\
         $C_\mathrm{hanging}$ & $52 \pm 1$  fJ/K \\
         $G_\mathrm{TES,hanging}$ & $655 \pm 15$  pW/K \\
         $M$ & $2.2 \pm 0.4$  \\
        \hline
        \hline
    \end{tabular}
\end{table}

\subsection{Nonlinearity and gain}

\begin{figure}
    \centering
    \includegraphics[width=1\linewidth]{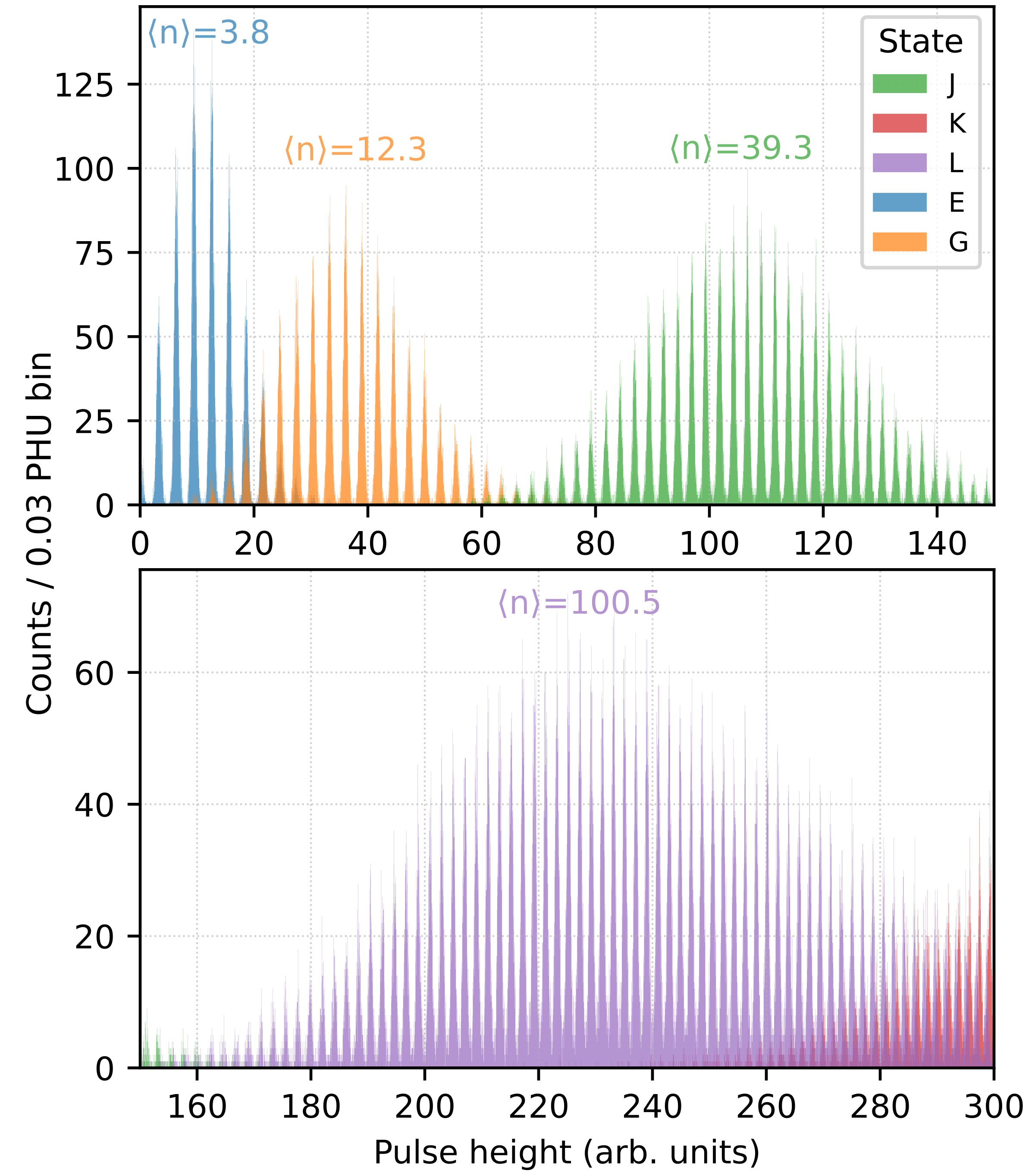}
    \caption{Pulse-height histograms for laser photons collected on Pixel 6 at different laser brightness settings. The average photon numbers $\langle n \rangle  $ for each Poisson-distributed group are shown above it.}
    \label{fig:PH-histogram}
\end{figure}

\begin{figure}
    \centering
    \includegraphics[width=1\linewidth]{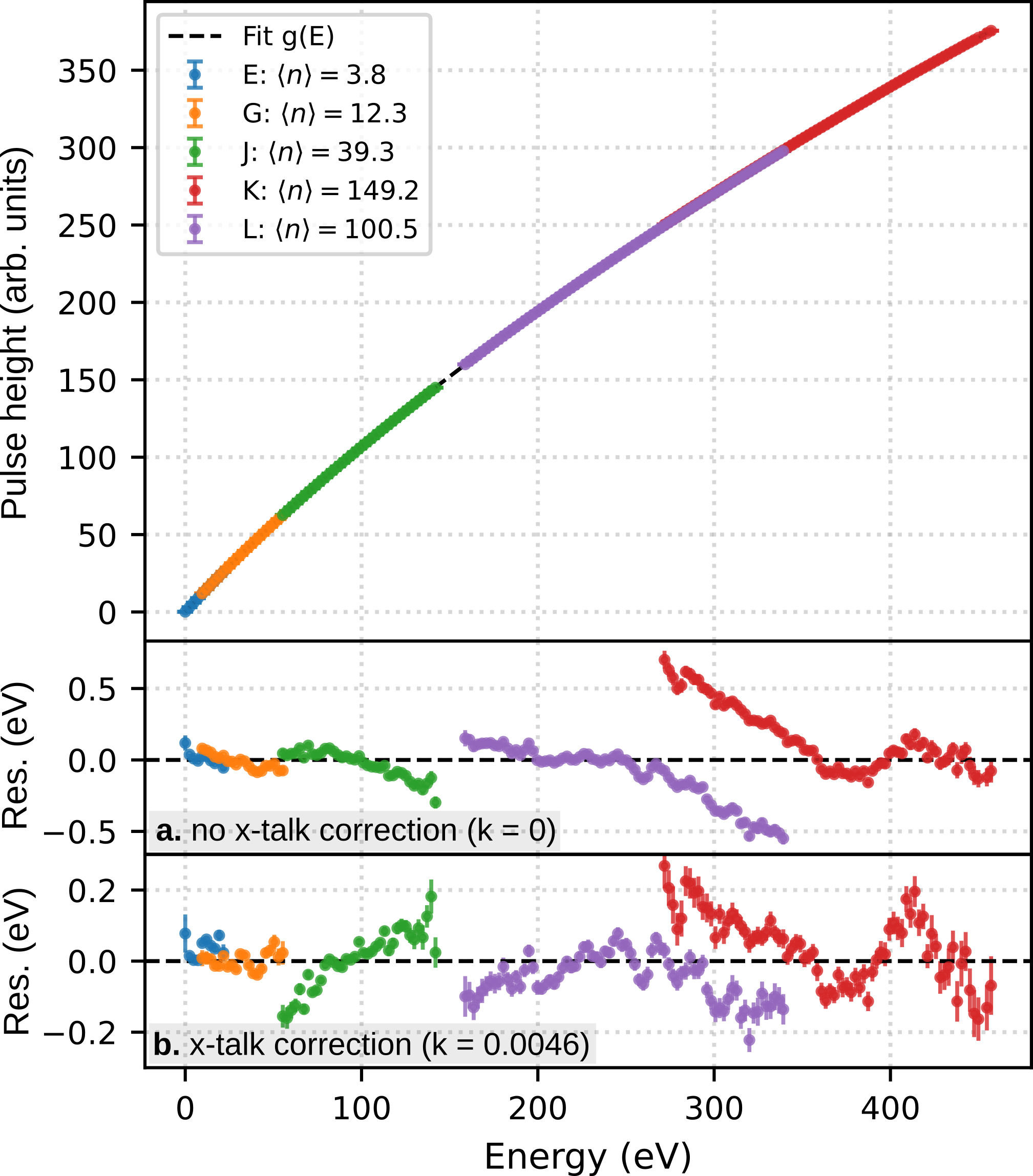}
    \caption{\textbf{a} A 7$^\mathrm{th}$-order polynomial fit to the centroids obtained from the pulse-height histogram vs. energy for Pixel 6. Residuals show evidence of thermal cross-talk arising from flood illumination.\textbf{b} Thermal-cross-talk corrected polynomial fit, with residuals shown. We find a total cross-talk level of about 0.5\%.
    }
    \label{fig:xtalk}
\end{figure}

We used \SI{2.4}{eV} photons produced by the pulsed laser to establish the detector's gain curve in the $<$~500~eV range and measure the achieved energy resolution in the small signal limit and at the carbon K-edge. The laser intensity was varied by adjusting the pulse-width (kept under \SI{100}{ns}) and by using a variable optical attenuator. Flood illumination from the laser produces Poisson-distributed photon combs on each pixel, visible in the pulse-height histograms shown for a single pixel in Fig.~\ref{fig:PH-histogram}. We fit each comb with a sum of Gaussians to precisely determine the centroid of each multi-photon peak. Plotting these centroids against the known incident energies, \(E_n = n E_{\gamma}\), defines the detector's gain curve (Fig.~\ref{fig:xtalk}).

This gain curve exhibits significant nonlinearity (about 27\% at carbon K-edge), which we initially model using a 7$^\text{th}$-order polynomial. While this provides a good global fit, the residuals show a step-like structure, consistent with thermal crosstalk. During flood illumination, photons absorbed by neighboring pixels generate heat that diffuses to the target pixel, slightly increasing its pulse height. This manifests in Fig.~\ref{fig:xtalk}(a) as offsets between photon combs of equal energy but different illumination levels (e.g., between the purple and red points). The crosstalk energy shift, \(\delta E_n\), is proportional to the average number of absorbed laser photons $\langle n \rangle$\cite{jaeckel2021calibration}. Therefore, we model the pulse height as a polynomial function (order $m$) of an effective energy:
\inserteq{\mathrm{PH} = f(E, k) := \sum_{i>0}^m a_i\left(E + k\langle n \rangle E_{\gamma}\right)^i,}{eq:xtalk2}
where $k$ quantifies the level of thermal crosstalk.

Our crosstalk correction algorithm proceeds in two steps. First, we fit the uncorrected pulse-height centroids as a function of energy to extract $k$ and the $a_i$ coefficients. We find the all-to-one thermal crosstalk $k = 4.6\times10^{-3}$. The contribution of the crosstalk to the pulse heights can then be written as: 
\inserteq{\Delta \mathrm{PH}(E, \langle n \rangle) = f(E,  k) - f(E, k=0).}{eq:xtalk3} 
Next, using Eqs. \ref{eq:xtalk2} and \ref{eq:xtalk3} we define a corrected pulse height:
\inserteq{\mathrm{PH}' = g(E, k') := \mathrm{PH} - s \Delta \mathrm{PH},}{}
where \(s\) is a scaling factor that is optimized using a numerical routine to yield $k' = 0$. This effectively separates the intrinsic nonlinearity, described by a final polynomial \(g(E)\), from thermal crosstalk. After correction, the residuals are reduced to the \SI{0.25}{eV} level (Fig.~\ref{fig:xtalk}(b)). This characterization supports our strategy to pair flood illumination with focused laser calibration~\cite{dean2025pulsed} for accurate mapping of crosstalk in kilo-pixel arrays. Further study will be required to isolate intrinsic chip-level cross-talk from external sources, such as leakage from the flood illumination aperture.

\subsection{Energy resolution}
\begin{figure}
            \centering
            \includegraphics[width=1\linewidth]{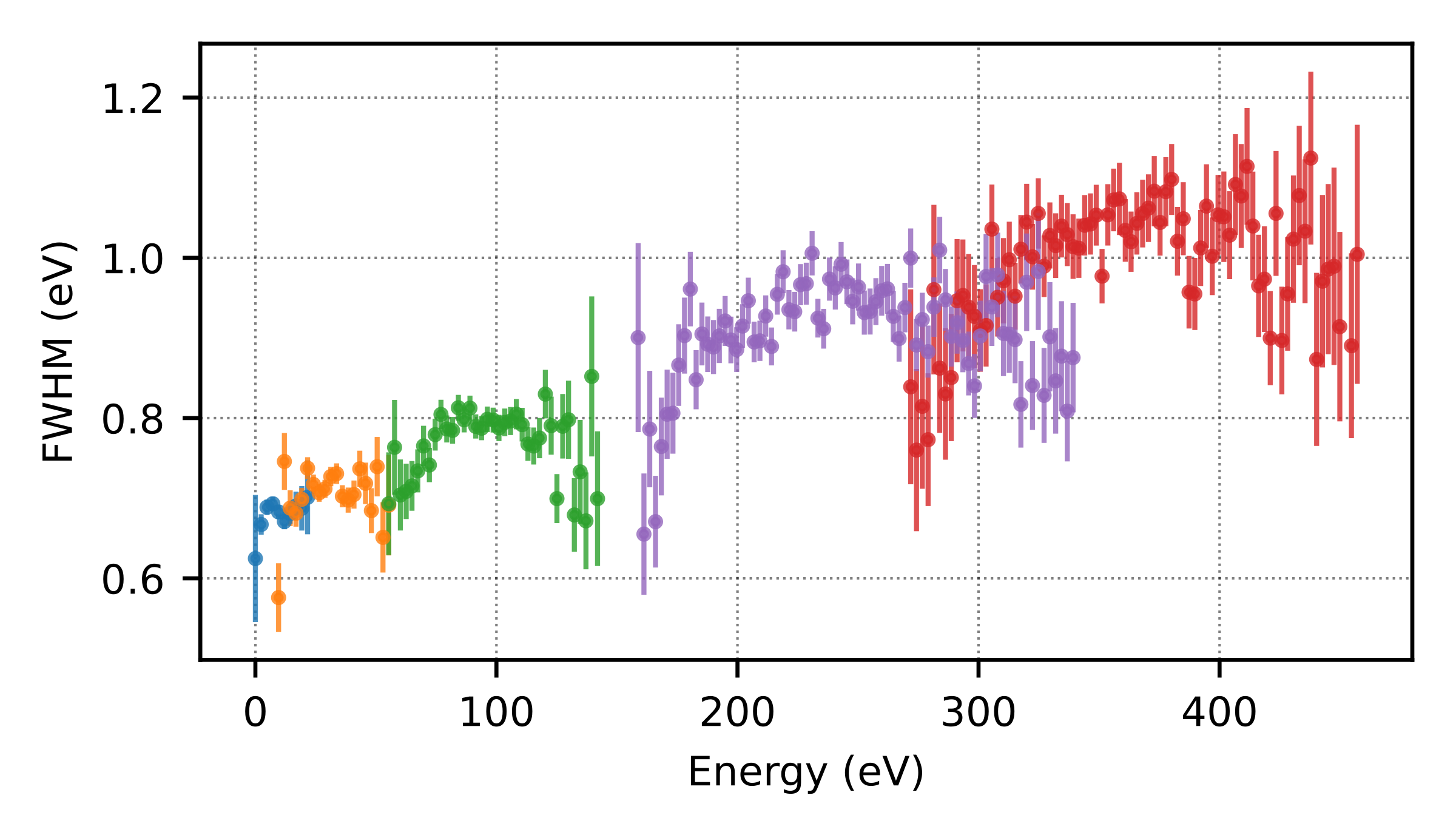}
            \caption{Energy resolution for Pixel 6, showing \SI{0.9}{eV} FWHM at the carbon K-edge at $\sim$ \SI{280}{eV}. The energy resolution has a square-root dependence in energy, which is expected to arise from downconversion phonon noise~\cite{kozorezov2006resolution}.}
            \label{fig:energy-resolution}
\end{figure}

The energy resolution of our detector is shown in Fig. \ref{fig:energy-resolution}. These devices, with \SI{250}{\micro m} square absorbers and a $T_{c} \approx$ \SI{53}{mK}, achieve \SI{0.9}{eV} FWHM at the carbon K-edge. Our electrothermal model is consistent with this measured performance; the measured resolution of \SI{0.65}{eV} at 2.4~eV is in good agreement with the model's prediction of \SI{0.58}{eV}.

This validated model serves as a useful tool for optimizing future pixel designs. It predicts that reducing the absorber size to \SI{200}{\micro m} square and lowering the critical temperature to \SI{30}{mK} would yield an energy resolution of 0.30 (0.23)~eV at \SI{300}{eV} for our sidecar (overhanging, assuming no hanging heat capacity) absorber pixels, which would satisfy our scientific requirements. A measured factor-of-three margin in the detector dynamic range provides flexibility to trade saturation energy for potential improvements in spectral resolution.

\section{Conclusion}
We have successfully fabricated and characterized prototype transition-edge sensor (TES) microcalorimeters on a flexible, all-silicon architecture. Despite the choice of a novel substrate, we find that the detector performance remains acceptable, establishing a critical baseline for future large-format soft X-ray spectrometers. A detailed electrothermal model parametrized by I-V curves and complex impedance measurements accurately predicts the measured energy resolution of our test devices. We validated our electrothermal model by comparing its predictions to photon-counting spectra from a pulsed optical laser, confirming its accuracy in the small-signal limit. We have also demonstrated that the thermal conductance of these pixels can be reliably predicted from device geometry and fundamental silicon material properties, showcasing the robustness and predictability of our SOI-based fabrication process.

The validated model is a powerful predictive tool for future designs. Using this model, we are designing a pixel with $T_\mathrm{c} \sim $ \SI{30}{mK} that is projected to achieve an energy resolution of \SI{0.3}{eV} at the carbon K-edge. This work provides the essential characterization and modeling foundation for this next step. This pixel and the planned array will enable a high-efficiency soft x-ray spectrometer to be installed on the NSLS-II beamline for the purpose of studying carbon catalysis.


\bibliographystyle{IEEEtran}
\bibliography{references}

\vfill

\end{document}